\documentclass[a4paper,fleqn]{cas-dc}

\usepackage[numbers]{natbib}
\usepackage{upgreek}
\def\tsc#1{\csdef{#1}{\textsc{\lowercase{#1}}\xspace}}
\tsc{WGM}
\tsc{QE}
\tsc{EP}
\tsc{PMS}
\tsc{BEC}
\tsc{DE}

\begin{document}
\let\WriteBookmarks\relax
\def\floatpagepagefraction{1}
\def\textpagefraction{.001}


\title [mode = title]{
Transport of aerosols and nanoparticles through respirators and masks
}


%
\author[1,2]{KV Chinmaya}

\cormark[1]

\fnmark[1]

\ead{chinmay@centerfornanodevices.com}



\affiliation[1]{organization={International Center for Nanodevices},
    addressline={TBI-InCeNSE, Indian Institute of Science Campus}, 
    city={Bengaluru},
    postcode={560012}, 
    state={Karnataka},
    country={India}}

\author[1,2]{Moumita Ghosh}[style=chinese]

\author[1]{G Mohan Rao}


\affiliation[2]{organization={Open Academic Research Council},
    city={Kolkata},
    postcode={712235}, 
    state={West Bengal},
    country={India}}



\author%
[1,2,4]
{Siddharth Ghosh}
\cormark[2]
\fnmark[1,2]
\ead{sg915@cam.ac.uk}

\affiliation[4]{organization={Department of Applied Mathematics \& Theoretical Physics, University of Cambridge},
    city={Cambridge},
    postcode={CB3 0WA}, 
    country={UK}}

\cortext[cor1]{Corresponding author}
\cortext[cor2]{Principal corresponding author}

\begin{abstract}
In several countries wearing multiple surgical masks or N95 respirators was mandatory in public during the COVID pandemic. 
Concerns arose about potential health issues caused by the unnatural breathing force required to maintain the breathing cycle when wearing these masks. 
In this study, we investigated the transportation and filtering mechanism of heterogeneous nanoparticles and viruses through surgical masks and N95 respirators. 
We conducted experiments \textit{in vitro} using aerosol spray paints containing nanoparticles and validated the findings \textit{in vivo} on a human volunteer. 
Scanning electron microscopy was employed to analyse the transportation and distribution of nanoparticles in different mask layers and on pristine silicon substrates placed on human skin. 
We provide analytical insights into the pressure distribution and fluid velocity profiles within the complex polymer network.
Remarkably, our results showed that both single surgical masks and N95 respirators demonstrated similar efficiency in filtering colloidal and jet-stream nanoparticles in the air. 
These comprehensive findings have significant implications for policymakers in defining regulations for airborne pandemics and air pollution control.
\end{abstract}

\begin{graphicalabstract}
\includegraphics[scale = 0.5]{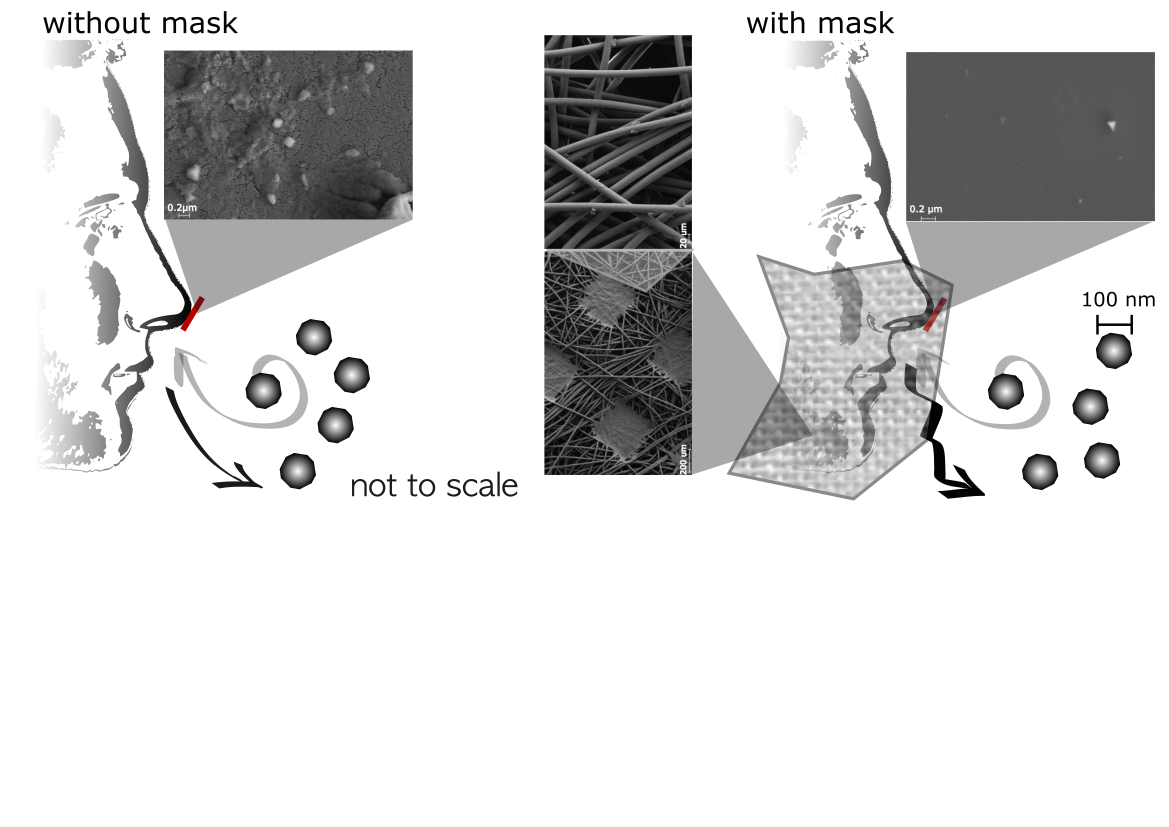}
\end{graphicalabstract}

\begin{highlights}
\item For the purpose of filtering colloidal and jet-stream nanoparticles in the air, we discovered that single surgical masks and N95 respirators give outcomes that are comparable. 
\item We examined the transport and distribution of nanoparticles on clean silicon substrates in humans and in various layers of masks and respirators using scanning electron microscopy.
\end{highlights}


\maketitle
\section{Introduction}

The outbreak of the COVID-19 pandemic compelled the widespread use of facial respirators and masks \cite{wikiCovid}. 
These tools play a crucial role in reducing the risk of viruses, pollutants, and other droplets entering the human body \cite{belkin1997evolution,cherrie2018effectiveness,aragaw2020surgical,prather2020reducing, worby2020face,betsch2020social,coclite2021face,liao2021technical}. 
Various types of protections, such as N95 respirators, cotton masks, surgical masks, and polymeric masks, are commercially available, each serving specific purposes \cite{li2015design,jiang2016air,wang2020mask,kodros2021quantifying}. 
Their filtration capability is influenced by factors like fiber type, manufacturing method, web structure, and fiber's cross-sectional shape. 
A comprehensive investigation is required to understand the transport mechanism of nanoparticles for optimal mask design. 
Three-ply surgical masks with a melt-blown material sandwiched between non-woven fabric are commonly used \cite{loeb2009surgical}, while N95 respirators typically consist of five layers of fabric \cite{juang2020N95}. 
During the pandemic, mask analyses focused on efficacy and filtration levels \cite{bundgaard2021effectiveness, eikenberry2020mask, brooks2021effectiveness,wang2021effective,lustig2020effectiveness,abbas2021cost,konda2020aerosol,gogoi2021nanometer}. 

Our research introduces a novel method to determine the effectiveness of nanoparticles adsorbed in N95 and surgical masks. 
Utilising scanning electron microscopy, we conducted an in-depth study of the internal network and architecture of masks. 
By analysing every layer of the mask, we gained insights into the filtration process. 
A human volunteer wore N95 respirators and surgical masks while analysing aerosol particles from spray paints and naturally occurring nanoparticles/viruses. 
To determine the number and form of nanoparticles in the filtered air, we examined the internal layers of various masks on pristine cleanroom-grade silicon. 
We conducted experiments on particle filtration by positioning the silicon wafer near the nose, where respiration occurs. 
The research highlights the importance of single surgical masks and N95 respirators, demonstrating that they yield comparable efficiency in filtering colloidal and jet-stream nanoparticles in the air. 
This finding holds significant implications for policymakers in developing regulations to combat airborne pandemics and address air pollution concerns effectively. 
The study sheds light on the critical role played by facial respirators and masks in respiratory protection and public health policies.

\begin{figure}[]
    \centering
    \includegraphics[width=0.5\textwidth]{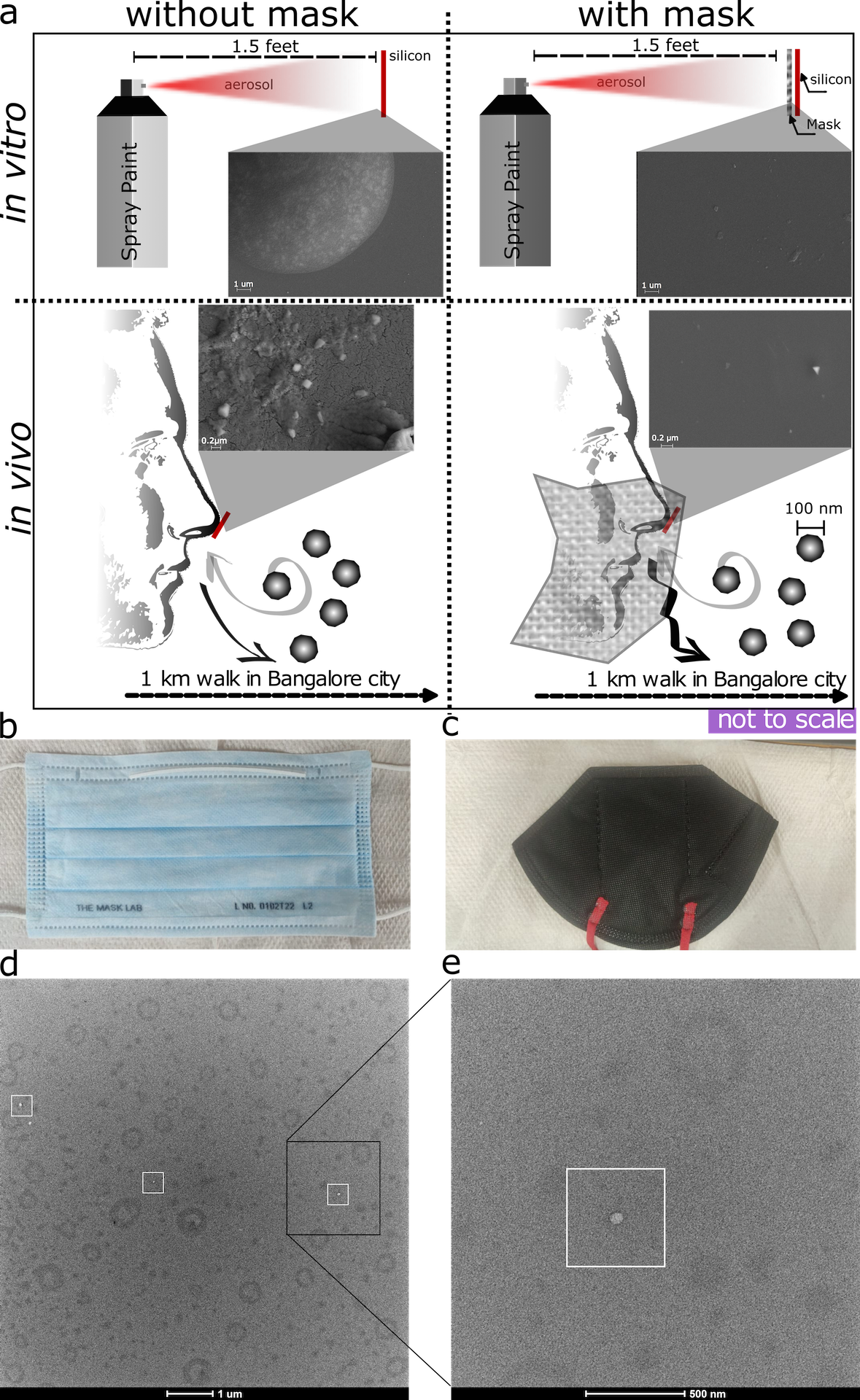}
    \caption{\textbf{Schematic representation of experimental setup of transport of molecules in the mask.} \textbf{a}, \textit{in vitro} -- a schematic diagram of the experiment which illustrates the nanoparticles entering the silicon wafer with and without a mask as a barrier in an open environment -- particles with their size and its visualization techniques. \textit{In vivo} -- a schematic diagram of the experiment which illustrates the particles which come from spray paints, are filtered out with three layers and finally trapped on the silicon wafer.
    \textbf{b}, surgical mask used for the experiment. 
    \textbf{c}, N95 respirator used in the experiment.
    \textbf{d}, transmission electron microscopy image of spray paint nanoparticle distribution shown in the white box at low concentration.
     \textbf{e}, magnified image of the spray paint nanoparticle with the size of 50 nm.}
    \label{figure:1}
\end{figure}

\section{Experimental setups}

The experimental setups for both \textit{in vitro} and \textit{in vivo} conditions are presented in Figure \ref{figure:1}\textbf{a}. 
This figure showcases the entry of nanoparticles onto the silicon wafer, with and without a mask as a barrier, in open environment particles and spray paint nanoparticles, along with their size and visualization techniques. 
In the \textit{in vitro} experiment, spray paint particles are filtered out through three layers and eventually trapped on the silicon wafer. 
The choice of spray paint is due to its vigorous inhalation-mimicking air pressure outflow from the jet. 
The \textit{in vivo} experiment, also illustrated in Figure \ref{figure:1}\textbf{a}, aims to understand the transportation of nanoparticles through masks and respirators in atmospheric air while a human breathes and walks over 1 km. 
The road map and distance travelled can be found in the supporting information in Figure 1.
Figures \ref{figure:1}\textbf{b} and \ref{figure:1}\textbf{c} display the surgical mask and N95 respirator used in the experiment, providing protection from airborne microorganisms when sneezing, coughing, or speaking during an airborne pandemic\cite{arumuru2020experimental}. The transmission electron microscopy analysis in Figures \ref{figure:1}\textbf{d} and \ref{figure:1}\textbf{e} confirms that the spray paint particles indeed contain nanoparticles.
To conduct the experiments, pristine silicon wafers were placed inside the masks on top of the nose to sample nanoparticles penetrating through surgical masks and N95 respirators.
The silicon wafer pieces were properly cleaned, following the cleaning protocol in the supporting information (Figure 2). 
The human volunteer, one of the authors, is a healthy 26-year-old male, weighing 73 kg, and measuring 179 cm in height. 
The volunteer had not been diagnosed with COVID-19 and received two vaccinations with a recombinant, replication-deficient chimpanzee adenovirus vector encoding the SARS-CoV-2 Spike (S) glycoprotein. 
The experiment was conducted without endangering the volunteer's life, with the research plan reviewed and supervised by the other authors.
For the experiments involving aerosol particles from spray paints, silicon wafer pieces measuring 0.5 cm $\times$ 0.5 cm were used at a distance of 1.5 feet from the N95 respirator and surgical masks. 
Asian Paints ezyCR8 Apcolite Enamel Paint Spray Black 200ml was used to produce nanoparticles/aerosol particles. 
In the \textit{in vivo} experiment, 0.5 cm $\times$ 0.5 cm silicon wafers were placed inside the masks to collect nanodroplets from the open environment, both with a mask as a barrier and without a mask. 
The initial experiment was carried out in a controlled environment within a class 1000 cleanroom.
To image the nanoparticles, a Zeiss ULTRA 55 scanning electron microscope (SEM) with 5 kV of energy and a 30 $\mu$m aperture was utilized. 
To enhance resolution, 20 kV of electron energy was used to image spray paint particles on the silicon wafer. 
Prior to imaging, a 10 nm gold thin film was sputter-coated on each layer of the mask and silicon wafer pieces using the Quorum 150 R Gold plasma sputtering system to mitigate e-beam-generated charge-induced imaging artifacts from insulating materials. 
For transmission electron microscopy analysis, a carbon-coated copper grid was used. 
The spray paint particles were diluted 10,000 times in pure isopropanol, and the diluted solution was drop cast onto a TEM copper grid for imaging after drying. 
The Talos L120C G2 Transmission Electron Microscope from Thermo Fisher Scientific, equipped with the 4K $\times$ 4K Thermo Scientific Ceta CMOS Camera, was used for this analysis. 
Low and high magnifications were employed to analyse the low-concentrated spray paint nanoparticles.




\section{Results}
\subsection{\textit{In vitro} study with spray paint}
\begin{figure*}[!htp]
    \centering
    \includegraphics[width=1\textwidth]{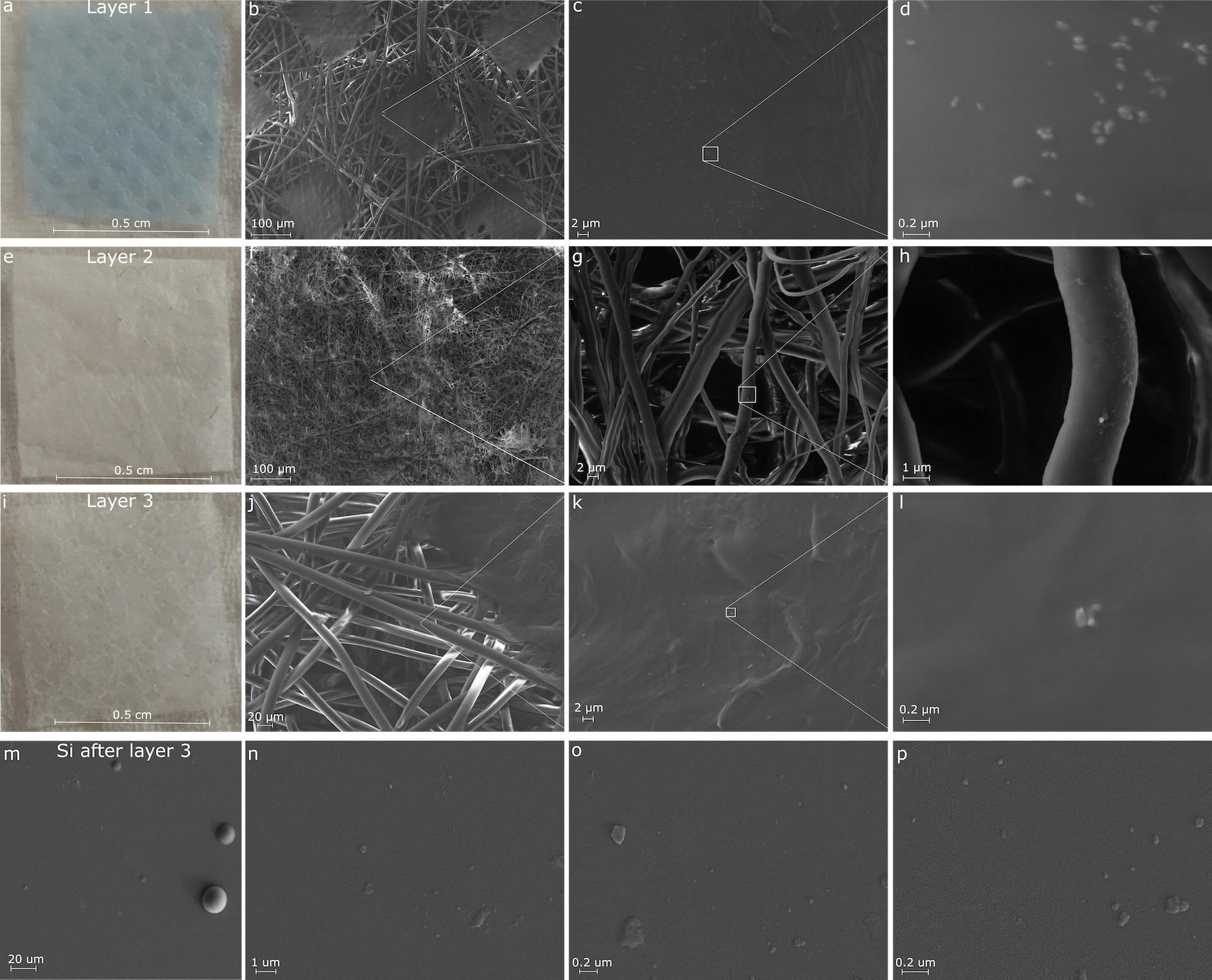}
    \caption{\textbf{Surgical mask} --- Digital and electron images of spray painted surgical mask layers and their relative particles.
    \textbf{a}, digital image of the first layer of surgical mask followed by electron magnified images in \textbf{b-d}.
    \textbf{e}, digital image of the second layer of the surgical mask followed by electron magnified images in \textbf{f-h}.
    \textbf{i}, digital image of the third layer of the surgical mask followed by electron magnified images in \textbf{j-l}.  
    \textbf{m}, spray paint particle deposited on the silicon wafers after it passes through the surgical mask and its corresponding images in \textbf{m-p}.}
    \label{figure:2}
\end{figure*}

The exterior layer, middle layer, and inner layer fibre structures are depicted in Figure \ref{figure:2}. Among these layers, the middle layer exhibits the most significant amount of fibre fabric, making it effective in preventing the entry of nanoparticles. In Figure \ref{figure:2}\textbf{a}, Figure \ref{figure:2}\textbf{e}, and Figure \ref{figure:2}\textbf{i}, the effects of aerosol particles from spray paints are shown on the surface of a surgical mask. The relative size of these particles ranges from 80 nm to 350 nm. The particles tend to aggregate, making individual particle visualization and focusing challenging.
The third layer of the surgical mask also contains particles, but its particle count is 26 times less than the second layer, as shown in the 200 nm scale SEM image in Figure \ref{figure:2}\textbf{j-l}. This reduction is attributed to the second layer, which forms a network of fibres providing filtration to the third layer.
In Figure \ref{figure:2}\textbf{m}, paint particles were deposited on silicon wafer pieces after passing through surgical masks, and their relative particles are shown in Figure \ref{figure:2}\textbf{n-p} at different areas and magnifications. 
To validate the results, we performed a control experiment with spray paints, observing spray paint particles directly deposited on a silicon wafer surface without any mask as a barrier. The magnified images of the particles ranging from 50 nm to 6 $\upmu$m are presented in supporting information Figure 4\textbf{a-d}.
For a more detailed analysis of the nanometer-sized particle distribution, we employed transmission electron microscopy, and the corresponding images are shown in supporting information Figure 5\textbf{a-f}.

\begin{figure*}[!htp]
    \centering
    \includegraphics[width=1\textwidth]{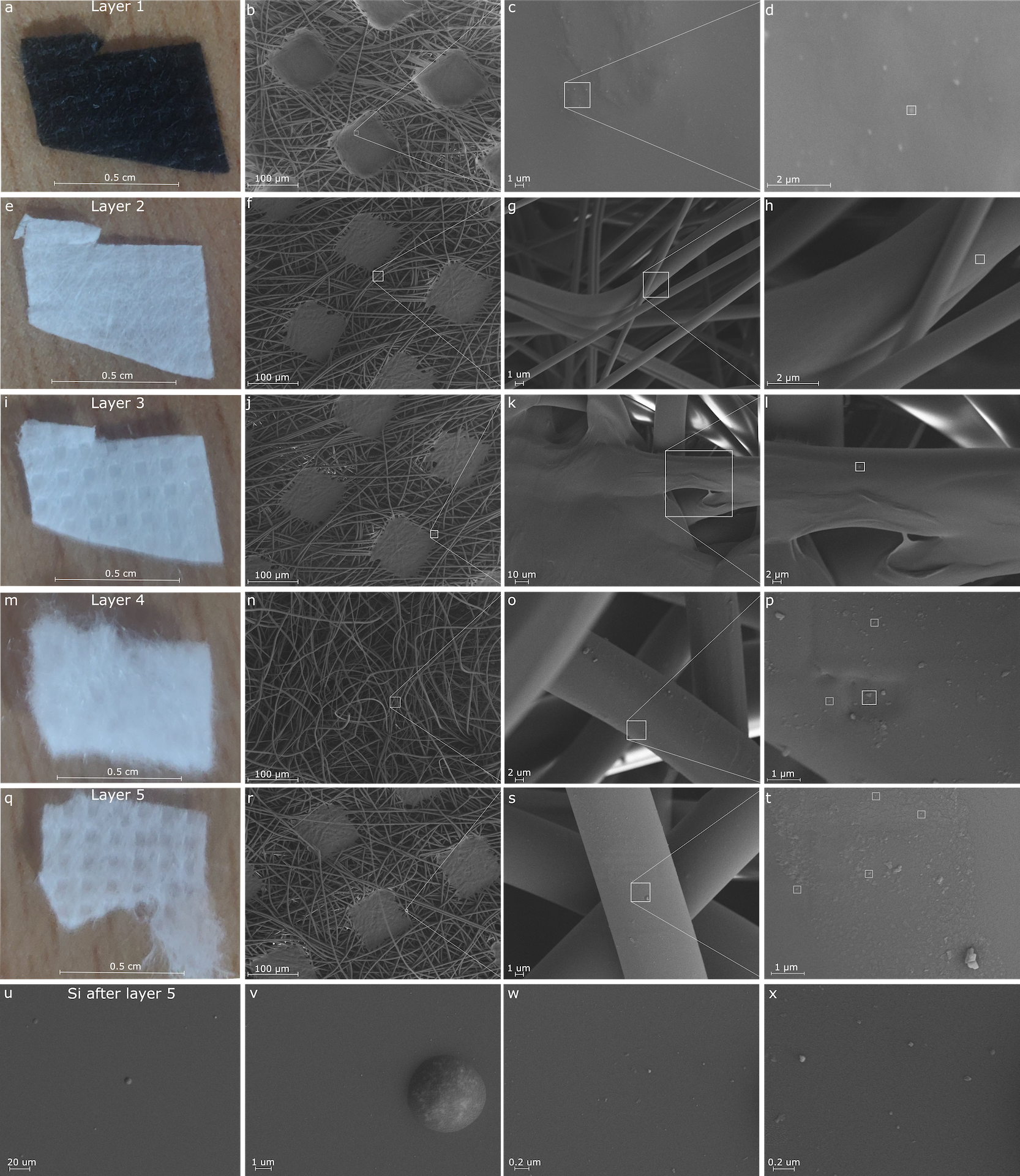}
    \caption{\textbf{N95 respirator} --- Digital and electron images spray painted N95 respirator layers and its relative particles.
    \textbf{a}, digital image of first layer of N95 respirator followed by magnified electron images in \textbf{b-d}.
    \textbf{e}, digital image of the second layer of the N95 respirator followed by magnified electron images in \textbf{f-h}.
    \textbf{i}, digital image of the third layer of the N95 respirator followed by magnified electron images in \textbf{j-l}.
    \textbf{m}, digital image of the fourth layer of the N95 respirator followed by magnified electron images in \textbf{n-p}.
    \textbf{q}, digital image of the fifth layer of the N95 respirator followed by magnified electron images in \textbf{r-t}.
    \textbf{u}, spray paint particle deposited on the silicon wafers after it passes through the N95 respirator and its corresponding images in \textbf{v-x}.}
    \label{figure:3}
    \end{figure*}

We also conducted experiments using N95 respirators, which consist of five layers of fabric. 
The aerosol particles from spray paint were dispersed in Figure \ref{figure:3}, showing a similar variation in relative particle size as observed in the surgical mask (Figure \ref{figure:3}\textbf{a-t}). 
The fifth layer of the N95 respirator (Figure \ref{figure:3}\textbf{t}) exhibited particles of a size similar to those seen at the end of the third layer in the surgical mask (Figure \ref{figure:2}\textbf{i}).
In the case of the N95 respirator, the particle count was reduced in the silicon wafer placed at the end layer of the mask. Figures \ref{figure:3}\textbf{u-x} display the paint particles deposited on the silicon wafer pieces after passing through the N95 respirator, along with their relative particle sizes. 
Similar particle sizes from spray paint were observed in both N95 respirators and surgical masks. 
In both mask types, a range of particles from 200 nm to 3000 nm was observed in the collected silicon wafer pieces. However, in the N95 respirator, the particle concentration was reduced, as evident in Figure \ref{figure:3}\textbf{u-x}.

\subsection{\textit{In vivo} transportation of nanoparticles through masks/respirator from open environment}
\begin{figure*}[!htp]
    \centering
    \includegraphics[width=0.95\textwidth]{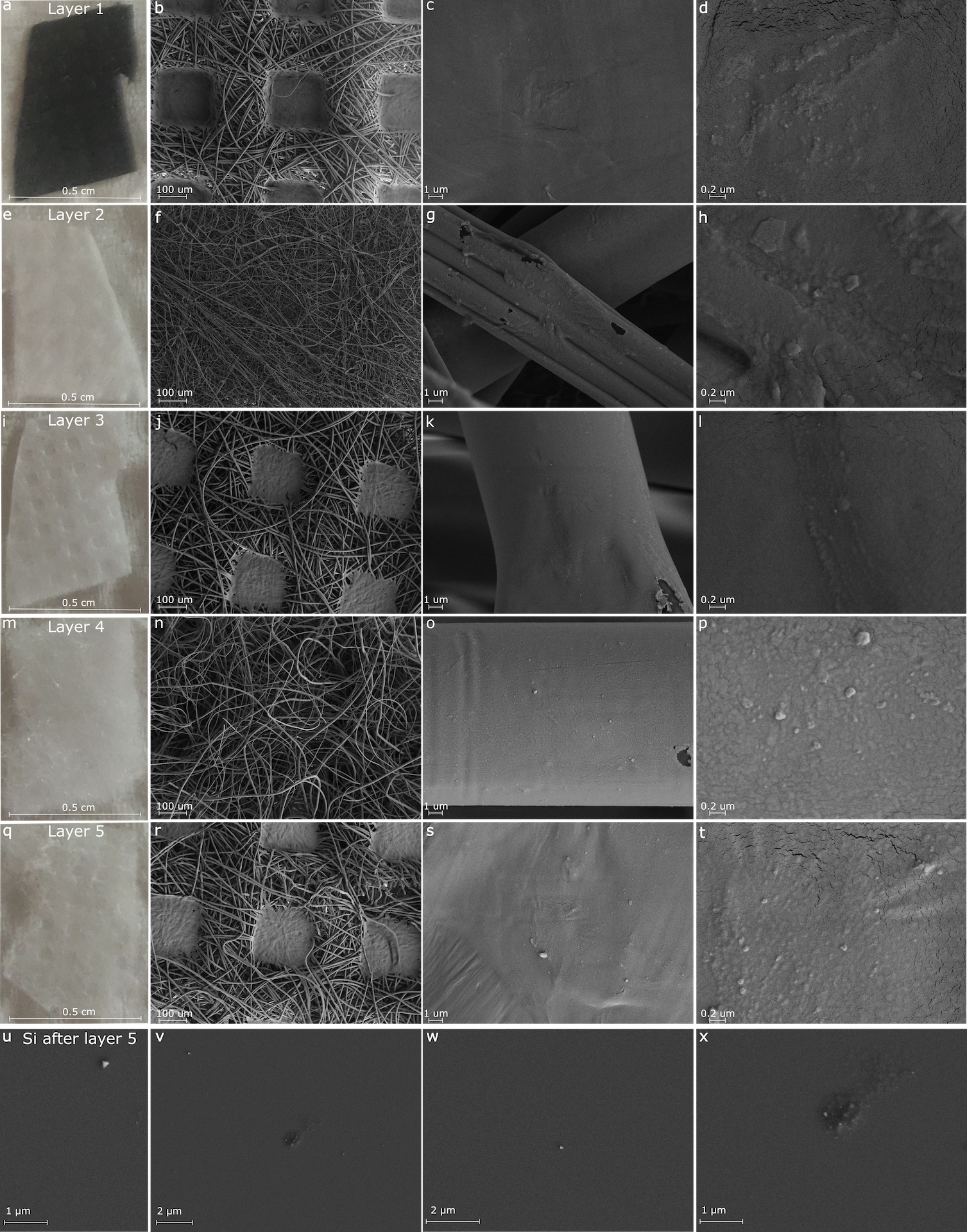}
    \caption{\textbf{N95 respirator} in an open environment which contains nano-droplets/particles from open air.
    \textbf{a}, digital image of first layer of N95 respirator followed by electron magnified images in \textbf{b-d}.
    \textbf{e}, an optical image of the second layer of the N95 respirator followed by electron magnified images in \textbf{f-h}.
    \textbf{i}, digital image of the third layer of the N95 respirator followed by electron magnified images in \textbf{j-l}.
    \textbf{m}, digital image of the fourth layer of the N95 respirator followed by electron magnified images in \textbf{n-p}.
    \textbf{q}, digital image of the fifth layer of the N95 respirator followed by electron magnified images in \textbf{r-t}.
    \textbf{u}, open environment particles deposited on the silicon wafers after it passes through the N95 respirator and its corresponding images in \textbf{v-x}.}
    \label{figure:4}
\end{figure*}

For the \textit{in vivo} transportation of nanoparticles through masks/respirators from an open environment, we conducted SEM imaging at different magnifications to analyse the particle distribution in the surgical mask and N95 respirator that could come from the environment.
In the supporting information Figure 3, we present surgical masks with all three layers in an open environment. 
Figure \ref{figure:4} showcases all five layers of the N95 respirator. 
Figure \ref{figure:4}\textbf{a-d} displays the first layer of the mask with aggregated nanoparticles on the surface. 
The second layer of the N95 respirator, shown in Figure \ref{figure:4}\textbf{e-h}, exhibits a fiber network designed to reduce particle counts.
In the next layer (Figure \ref{figure:4}\textbf{i-l}), the particle counts were further reduced, with particle sizes ranging from 60 nm to 190 nm, similar to the third layer of the surgical mask. The fourth and fifth layers are shown in Figure \ref{figure:4}\textbf{m-p} and Figure \ref{figure:4}\textbf{q-t}, respectively, with particle sizes ranging from 80 nm to 180 nm.
Figure \ref{figure:4}\textbf{u-x} shows the silicon wafer placed under the N95 respirator. 
In both cases, supporting information Figure 3\textbf{m-p} and Figure \ref{figure:4}\textbf{u-x}, the particle size remains consistent. 
Thus, N95 respirators and surgical masks demonstrate similar efficacy in blocking external particles from the environment.
We also performed a control experiment in an open environment, capturing particles on a silicon wafer, and collecting nanodroplets/particles in pristine silicon wafer pieces. Supporting information Figure 4\textbf{e-h} presents snapshots of nano-droplets on clean silicon wafer parts recorded in an open environment without a mask.

\begin{figure*}[!htp]
\centering
    \includegraphics[width=0.9\textwidth]{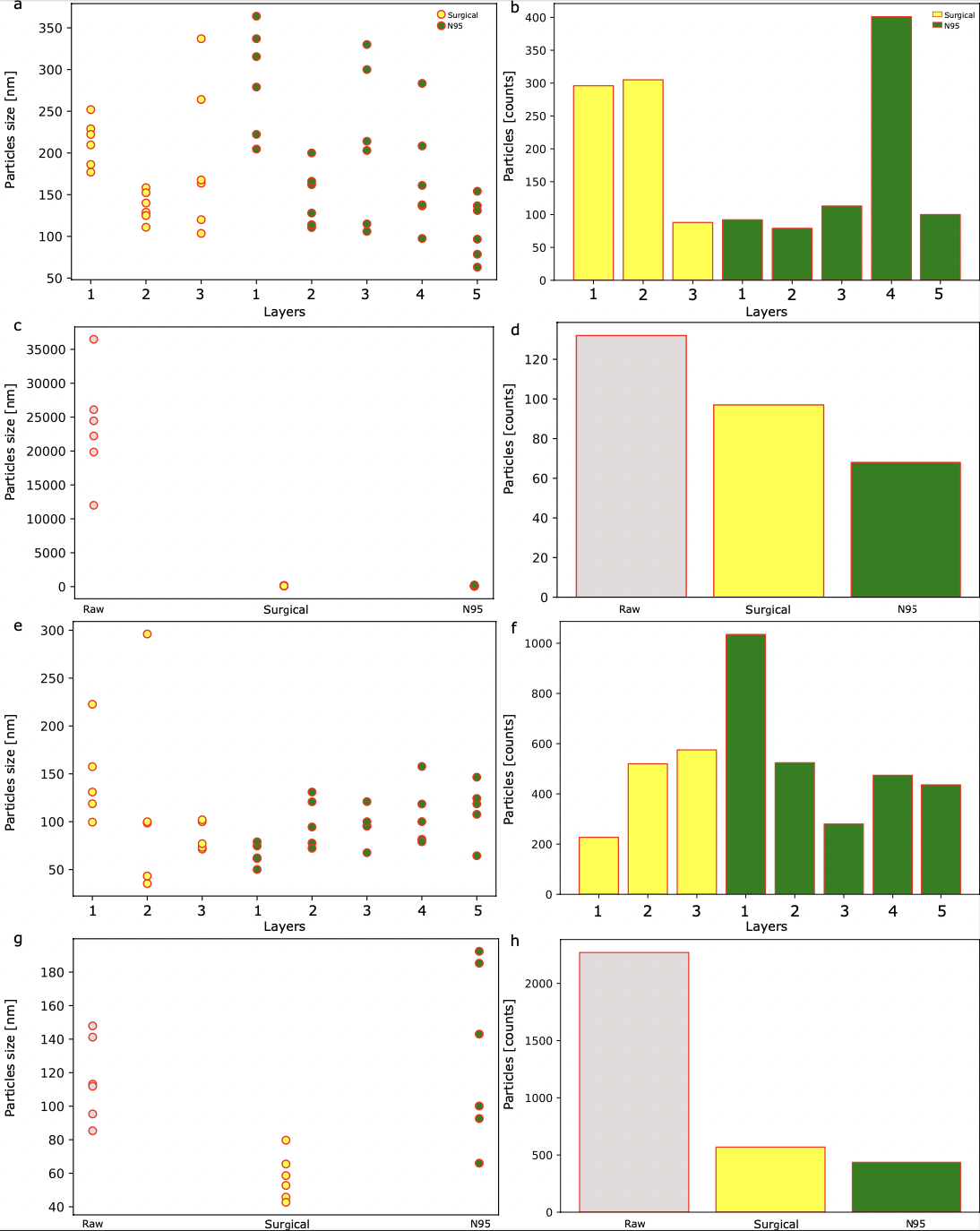}
     \caption{\textbf{Surgical mask and N95 respirator} ---
     Quantitative investigation of spray paint and open environment particle distribution on silicon wafers after passing through surgical masks and N95 respirators.
    \textbf{a}, particle size in different layers of a surgical mask and an N95 respirator for \textit{in vitro} study with spray paint.
    \textbf{b}, particle counts different layers of a surgical mask and an N95 respirator for \textit{in vitro} study with spray paint.
    \textbf{c}, particle size on silicon; raw refers to no filter was used. 
    \textbf{d}, particle counts on silicon; raw refers to no filter was used.
    \textbf{e}, particle size in different layers of a surgical mask and an N95 respirator for \textit{in vivo} study with air filtering with human assistance.
    \textbf{f}, particle counts in different layers of a surgical mask and an N95 respirator for \textit{in vivo} study with air filtering with human assistance.
    \textbf{g}, particles size on silicon wafers after it passes through a surgical mask and N95 respirator along with control in an open environment for \textit{in vivo} study with air filtering with human assistance.
    \textbf{h}, particles count on silicon wafers after it passes through the surgical mask and N95 respirator along with control in an open environment for \textit{in vivo} study.}
   
    \label{figure:5}
\end{figure*}

To provide a quantitative analysis of our findings, we performed extensive measurements from spray paint particles to nanoparticles in an open environment. Figure \ref{figure:5} presents the results of these measurements.
In Figure \ref{figure:5}\textbf{a} and Figure \ref{figure:5}\textbf{b}, we display the particle sizes found in each layer of both the surgical mask and the N95 respirator, as well as the number of particles found in each layer. For the first layer of the surgical mask, the particle size ranged from 175 nm to 250 nm, followed by 100 nm to 160 nm in the second layer, and 100 nm to 320 nm in the third layer. Similarly, particle variations were observed for the N95 respirator: 200 nm to 360 nm in the first layer, 100 nm to 200 nm in the second layer, 100 nm to 320 nm in the third layer, 100 nm to 290 nm in the fourth layer, and 60 nm to 150 nm in the fifth layer. In Figure \ref{figure:5}\textbf{b}, we illustrate the particle counts in each layer, showing that both the surgical mask and the N95 respirator have approximately 50 particles per pixel size of 1024 $\times$ 768 at their last layer.
In Figure \ref{figure:5}\textbf{c} and Figure \ref{figure:5}\textbf{d}, we demonstrate the particle size and counts on the silicon wafer piece with and without a mask as a barrier, where the spray-painted particles were deposited. The size of the raw silicon wafer ranged from 10500 nm to 40000 nm, but in both the surgical mask and the N95 respirator, the particle size was reduced, and the particle counts decreased significantly. Approximately 100 particles were found in the surgical mask, whereas around 70 particles were found in the N95 respirator.
We also conducted experiments in an open environment, and the quantitative results are shown in Figure \ref{figure:5}\textbf{e}. For the first layer of the surgical mask, the particle size ranged from 100 nm to 230 nm, followed by 20 nm to 290 nm in the second layer, and 70 nm to 100 nm in the third layer. Similarly, for the N95 respirator, the particle sizes were 50 nm to 70 nm in the first layer, followed by 70 nm to 150 nm in the second layer, 70 nm to 140 nm in the third layer, 70 nm to 160 nm in the fourth layer, and 60 nm to 140 nm in the fifth layer. Figure \ref{figure:5}\textbf{f} shows the particle count distribution in each layer of both surgical masks and N95 respirators.
Figure \ref{figure:5}\textbf{g} presents the particle size distribution and the number of particles on the silicon wafer with and without surgical masks and N95 respirators. The particle size of the raw silicon wafer ranged from 80 nm to 150 nm. In surgical masks and N95 respirators, the particle counts were significantly reduced, but the particle sizes remained relatively consistent.
Additionally, we have shown the reproducibility curve of nanoparticle size entering through masks and without masks using surgical masks in an open environment in SI Figure \textbf{6}. The curve indicates particle sizes ranging from 50 nm to 30000 nm without a surgical mask, 50 nm to 5000 nm with a mask, and 50 nm to 4000 nm for spray paint particles with a surgical mask, respectively.

\clearpage

\newpage

\section{Discussion}
Here, we discuss the nanoparticle transport process through a complex network of masks and respirators using the Brinkman equation \cite{durlofsky1987analysis}.
The Brinkman equation can be expressed in vector form as:
\begin{equation}
    \rho \left( \frac{{\partial \mathbf{u}}}{{\partial t}} + \mathbf{u} \cdot \nabla \mathbf{u} \right) = -\nabla p + \eta \nabla^2 \mathbf{u} - \frac{\eta}{\kappa} \mathbf{u} + \rho \mathbf{g}
\end{equation}
where:
 \( \rho \): Fluid density,
 \( \mathbf{u} \): fluid velocity vector
 \( p \): fluid pressure,
 \( \eta \): Dynamic viscosity of the fluid, 
 \( \frac{{\partial \mathbf{u}}}{{\partial t}} \): time derivative of velocity, 
 \( \nabla^2 \mathbf{u} \): Laplacian of fluid velocity, 
 \( \kappa \): permeability of the porous medium, 
 \( \mathbf{g} \): acceleration due to gravity.
The pressure gradient \( \nabla p \) acts as the driving force for fluid flow through the porous polymer network. In the context of nanoparticle transport, the pressure gradient originates from the difference in pressure between the external environment and the interior of the mask or respirator.
The term 
 \( \rho \left( \frac{{\partial \mathbf{u}}}{{\partial t}} + \mathbf{u} \cdot \nabla \mathbf{u} \right) \) 
represents the momentum transfer due to fluid flow. 
This term captures the acceleration and advection of the fluid carrying nanoparticles as it traverses the complex network.
The term \( \eta \nabla^2 \mathbf{u} \) represents the diffusion of momentum due to viscous effects. 
In the context of nanoparticle transport, it characterizes the dissipation of energy within the fluid as it interacts with the polymer network. 
Nanoparticles experience drag due to these viscous interactions.
The term \( \frac{\eta}{\kappa} \mathbf{u} \) represents the flow resistance within the porous medium. 
It accounts for the interaction between the fluid and the polymer network. 
A higher permeability \( \kappa \) allows for smoother fluid flow, while a lower permeability increases flow resistance, influencing nanoparticle transport.
The term \( \rho \mathbf{g} \) represents the acceleration of the fluid due to gravity. 
In the context of nanoparticle transport, inertial effects become significant for larger nanoparticles or higher flow rates. 
They contribute to particle advection, influencing particle trajectories.
The pressure gradient \( \nabla p \) and fluid velocity \( \mathbf{u} \) influence the pressure distribution within the polymer network. 
As nanoparticles navigate through tortuous paths, they  experience deposition  onto the polymer fibers (as shown earlier under SEM) or pores due to differences in pressure and fluid flow.
At the end, the nanoparticles may or may not exit the polymer network, completing their transport journey. 
The interplay of pressure gradients, viscous effects, inertial forces, and permeability determines the time it takes for nanoparticles to traverse the network.
Incorporating the Brinkman equation into the description provides a mathematical framework to understand the intricate interactions between fluid flow, pressure gradients, viscous effects, and permeability in the context of nanoparticle transport through complex polymer networks. 
This approach enables a deeper exploration of the physical mechanisms governing filtration and particle deposition within masks and respirators.
\begin{figure}[]
    \centering
    \includegraphics[width=0.5\textwidth]{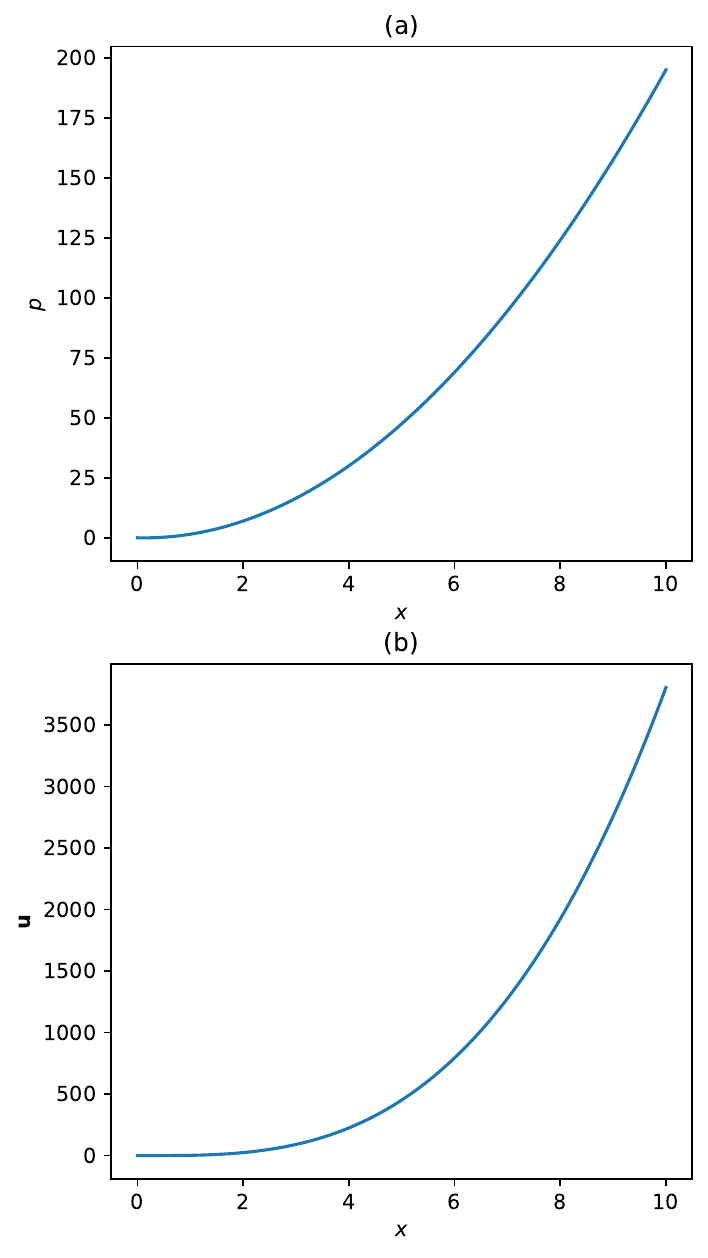}
    \caption{Analytical solutions for (a) $p$ and (b) $\mathbf{u}$ profile with respect $x$ using the simplified expressions of the Brinkman equation (5) representing nanoparticles' transport in complex polymer network of mask.}
    \label{fig:6}
\end{figure}
\begin{figure}[]
    \centering
    \includegraphics[width=0.5\textwidth]{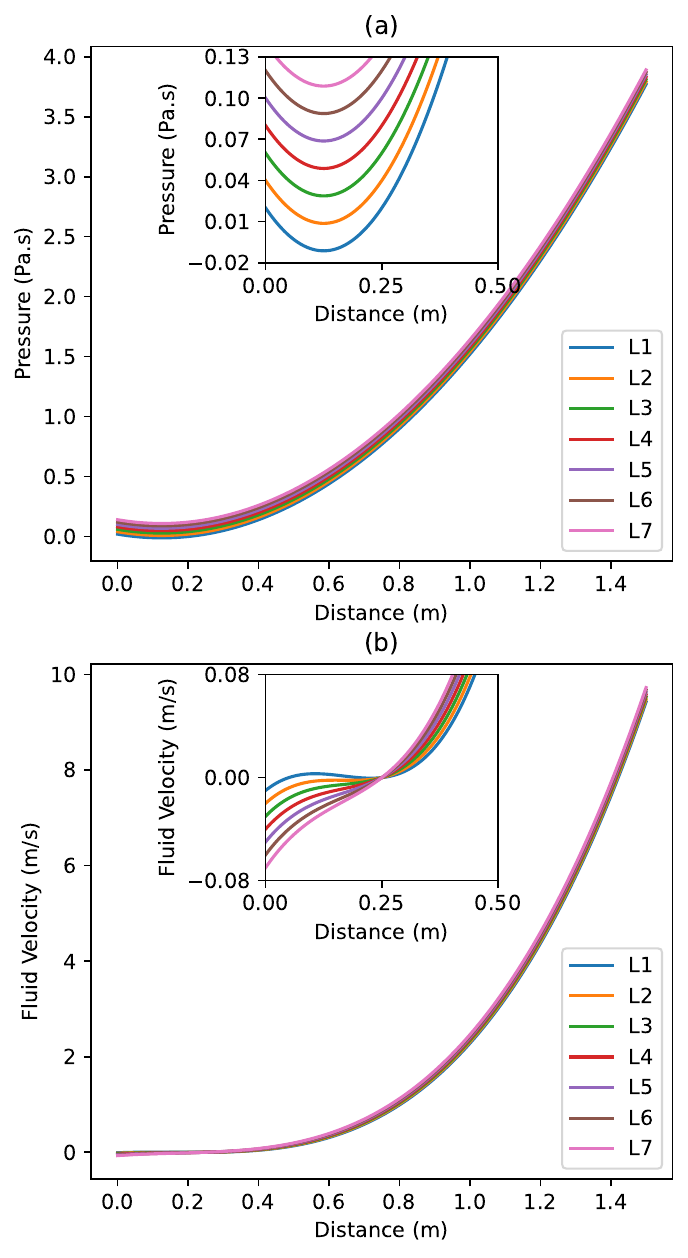}
    \caption{Change of (a) fluid pressure and (b) velocity profile with varying numbers of layers in the mask.}
    \label{fig:7}
\end{figure}
\subsection{Analytical Insights into Nanoparticle Transport through Polymer Networks}
Plotting the behaviours described by the Brinkman equation analytically is quite challenging due to the complex nature of the equation and the intricate interactions involved in nanoparticle transport through a polymer network \cite{diamant2015response, morozov2022mobility}. 
The Brinkman equation is a partial differential equation that involves multiple variables and terms, making it difficult to derive simple analytical solutions.
However, we attempted to simplify the situation by considering specific scenarios or simplified geometries. 
We consider a simplified case where we focus on the fluid flow and pressure distribution within a porous medium, without explicitly including inertial effects and nanoparticle dynamics. We'll focus on understanding the pressure distribution and fluid velocity within the polymer network.

Let's consider steady-state flow (no time derivative terms) and neglect the inertial and viscous terms for simplicity. This simplification results in a simplified form of the Brinkman equation:
\begin{equation}
    -\nabla p + \frac{\eta}{\kappa} \mathbf{u} = 0
\end{equation}
where
 \(\nabla p\) is  pressure gradient
.
In this simplified scenario, we can try to solve for the pressure distribution and fluid velocity analytically. Let's assume a 1D flow through a porous medium, where the porous medium has a linear permeability profile along the x-direction:
\begin{equation}
    \kappa(x) = \kappa_0 + \alpha x
\end{equation}
where
 \(\kappa_0\) is constant permeability term and
 \(\alpha\) is permeability gradient.
%
%
By solving the simplified Brinkman equation, we can obtain analytical expressions for pressure and fluid velocity profiles. However, due to the complexity of the equation, the resulting expressions might involve special functions or numerical integration.
The pressure distribution along the x-direction can be obtained by integrating the equation:
\begin{equation}
    -\int (\nabla p) dx + \frac{\eta}{\kappa_0} \int \mathbf{u} dx + \frac{\eta}{\alpha} \int x \mathbf{u} dx = C
\end{equation}
%
Assuming a steady flow (no time derivatives), the fluid velocity profile can be expressed as:
\begin{equation}
    \mathbf{u}(x) = -\frac{\alpha}{\eta} \int (\nabla p) dx + \frac{\kappa_0}{\eta} \int (\nabla p) x dx + A
\end{equation}
In these expressions, \(C\) and \(A\) are constants of integration that depend on boundary conditions.
The above solutions are simplified. 
They may not capture the full complexity of nanoparticle transport. 
Analytical solutions for a general case will involve inertial effects, nanoparticle dynamics, and complex geometries are likely to be much more complex as shown in our experiments. 
It may require numerical methods or simulations for accurate representation. 
We numerically solve and plot (Figure \ref{fig:6}) the simplified pressure distribution and fluid velocity profile using Python. 
It is meant for illustrative purposes and assumes a simple linear permeability profile. 
In our algorithm, we defined the constants (dynamic viscosity, permeability terms) and a linear permeability profile.
We then calculate the analytical solutions for pressure distribution and fluid velocity profile using the simplified expressions provided earlier. 
Later, we also introduced the effect multiple layers. 
We modified earlier algorithm to consider the number of layers in the mask and analyse how the pressure distribution and fluid velocity profile change with different layers. 
We have added a loop that iterates over different numbers of layers in the mask. 
For each number of layers, it calculates and plots the fluid pressure profile and velocity profile (Figure \ref{fig:7}).
The resulting plot shows how the fluid velocity profile changes with varying numbers of layers in the mask.
As we notice in our experiment, in reality, solving the Brinkman equation for complex scenarios would require involved numerical methods.
To accurately plot and analyse nanoparticle transport behaviours, numerical simulations or computational fluid dynamics techniques can be used afterwards to capture the intricate interactions and provide detailed insights into the transport process within a complex polymer network.

\section{Conclusion}
Our research has shown that aerosol particles generated by spray paint closely resemble those emitted during coughing, sneezing, or speaking by infected individuals. 
By utilising readily available fabric materials, both the general public and healthcare professionals can effectively reduce the risk of virus transmission through aerosols. 
Our study reveals that certain multi-layered mask designs offer filtration and adsorption capabilities on par with, or even superior to, the widely used five-layer N95 respirators.
Through comprehensive \textit{in vitro} experiments using aerosol spray paints containing nanoparticles, as well as \textit{in vivo} trials on a human volunteer in an open environment, we have demonstrated the efficacy of single surgical masks and N95 respirators in reducing particle counts. 
In both scenarios, comparable protection against external particles and aerosols was observed for both masks and respirators.
The repeatability of our data and multiple experiments further corroborate the effectiveness of N95 respirators in reducing particle transmission compared to surgical masks. 
Nevertheless, the findings indicate that both types of masks provide nearly comparable protection against nanoparticles and environmental pollutants.
This valuable insight into mask effectiveness will aid policymakers in establishing regulations and guidelines for future pandemics and pollution control measures. 
By promoting the widespread use of masks, we can significantly enhance public health and safety, mitigating the risks posed by aerosol-borne viruses and pollutants.

\section*{Acknowledgements} 

The authors thank the Honeywell CSR fund to the International Center for Nano Devices for funding.
The authors thank Dr. Samit Chakrabarty from Leeds University and Open Academic Research Council for advising us on human data collection.
The authors thank Dr. Somnath Dutta from Indian Institute of Science, Bengaluru for providing facilities to do experiment. 
The authors thank to Electron Microscopy Facility, Division of Biological Sciences, Indian Institute of Science, Bengaluru (supported by DBT-IISc Partnership Program Phase-II) for helping in transmission electron microscopy imaging.
The portion of characterisation work was supported by INUP-i2i at IISc Bangalore funded by Ministry of Electronics and Information Technology (MeitY). 
The work was performed using Facilities at CeNSE, Indian Institute of Science, Bangalore, funded by Ministry of Education (MoE), Ministry of Electronics and Information Technology (MeitY), and Nanomission, Department of Science and Technology (DST), Government of India. 

\section*{Author contributions statement}

SG proposed the problem.
KVC and SG conducted the experiments. 
KVC and SG analysed the results and wrote the manuscript. 
MG provided critical guidance on manuscript structuring. 
GMR supervised the research. 
SG designed and supervised the research. 
All authors reviewed the manuscript.

\section*{Competing interests}
The authors declare no conflict of interest.
\printcredits

\section*{Declaration of generative AI and AI-assisted technologies in the writing process}
During the preparation of this work the authors used ChatGPT 3.5 in order to polish the language. 
After using this tool/service, the authors reviewed and edited the content as needed and take full responsibility for the content of the publication.

\bibliographystyle{elsarticle-num}
\bibliography{ref}
\clearpage





\newpage

\section*{Supporting Information}
\begin{figure*}[]
    \includegraphics[width=1\textwidth]{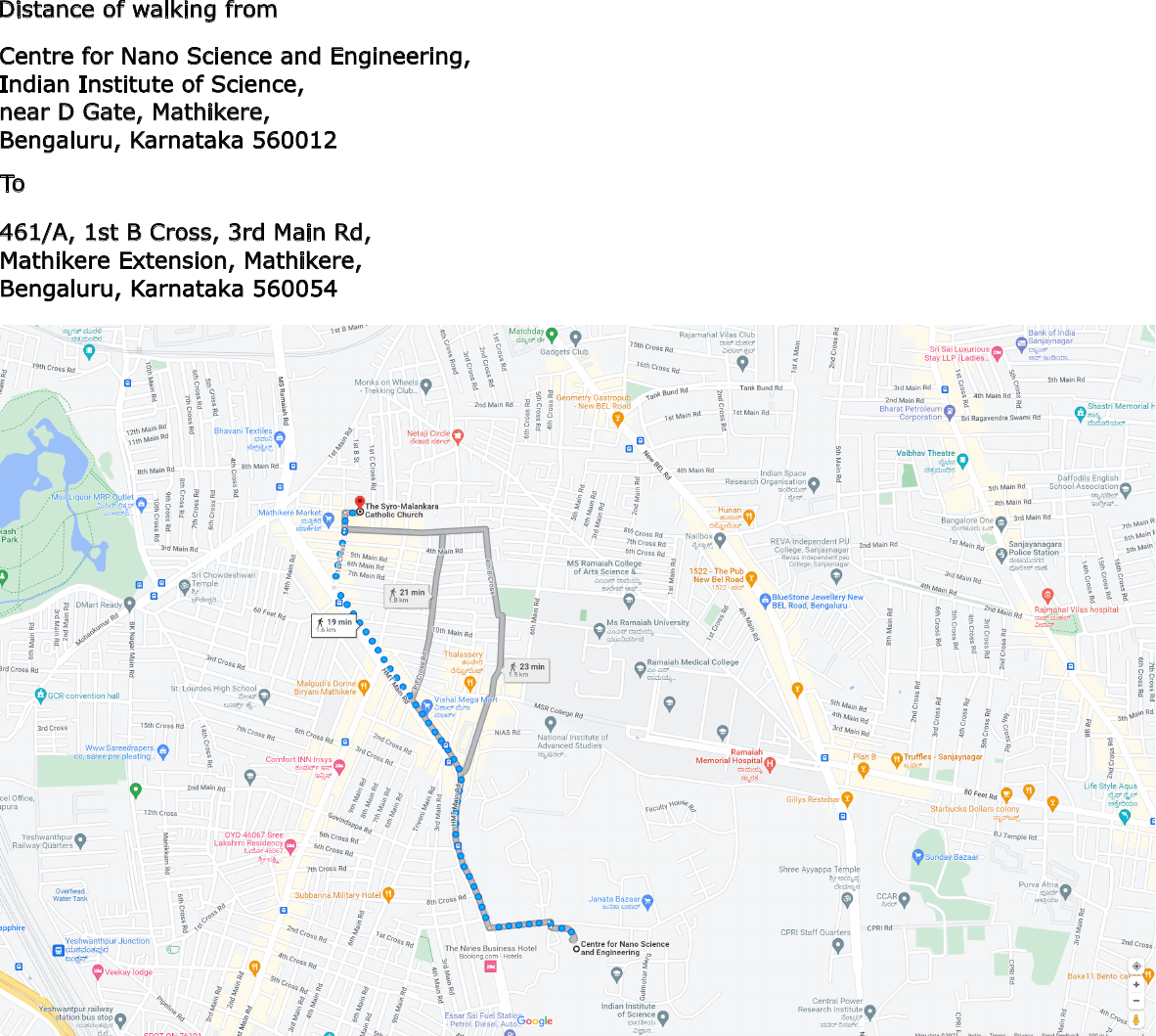}
    \caption{
    The Google map view -- the human volunteer has walked from Centre for Nano Science and Engineering, Indian Institute of Science, Bangalore 560 012, Karnataka, India to \#461/A, 1st B cross, 2nd Main road, Mathikere for the in vivo experiment. 
    The approximate distance is around 1.6 km}
    \label{SI Figure:1}
\end{figure*}

\begin{figure}[htp]
    \centering
    \includegraphics[width=0.5\textwidth]{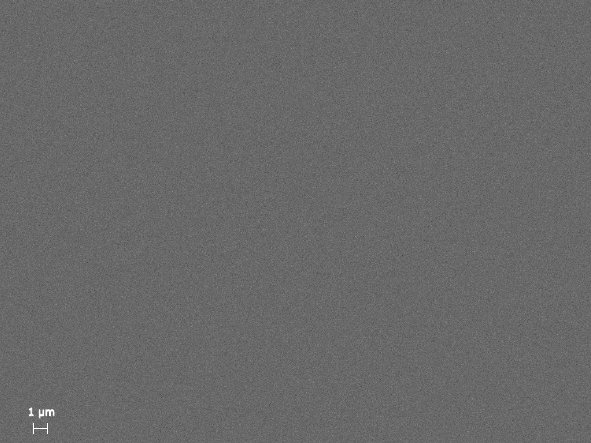}
    \caption{Control sample of silicon --- silicon wafer used before performing experiment.}
    \label{SI Figure:2}
\end{figure}

In Figure \ref{SI Figure:1}, we show the Google map view of the distance walked to perform \textit{in vitro} and \textit{in vivo} experiment.
The raw silicon piece after cleaning is shown in Figure \ref{SI Figure:2}.
The pristine silicon wafer was cleaned in class 1000 and class 100  cleanroom using piranha solution and solvents such as acetone and isopropyl alcohol. 
Silicon wafers pieces were ultrasonicated for 15 minutes in acetone, washed with de-ionised water, and nitrogen blown. 
After acetone cleaning, the silicon wafers pieces were again cleaned with isopropyl alcohol for 15 minutes, washed with de-ionised water, and nitrogen was blown. 
After this step, we baked the silicon wafers pieces at 110 degrees Celsius for 10 minutes. 
This will help to evaporate the excess of water droplets which was present in the wafer.

\begin{figure*}[htp]
    \centering
    \includegraphics[width=1\textwidth]{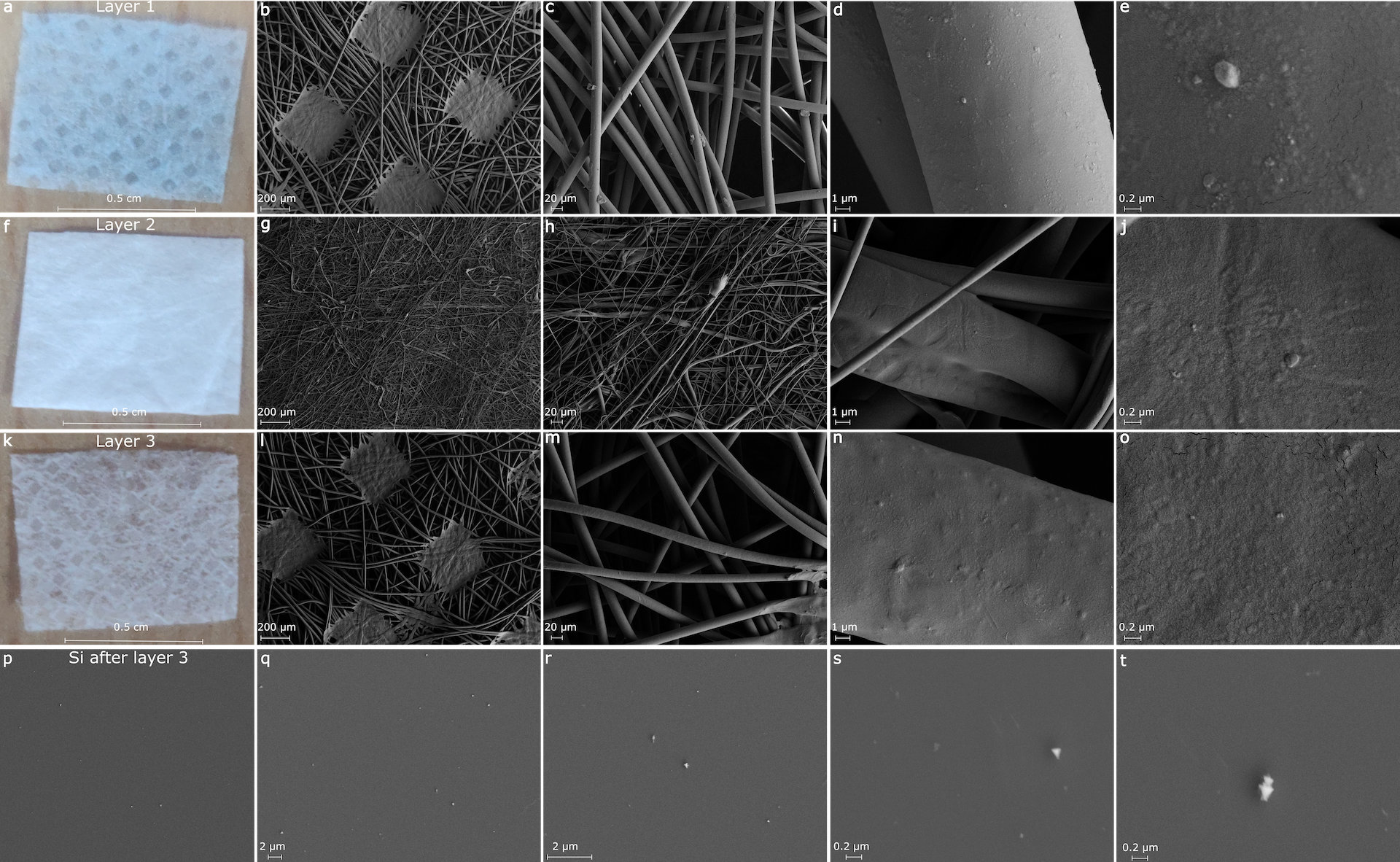}
     \caption{Surgical mask in an open environment which contains nano-droplets/particles from open air. 
    \textbf{a}, digital image of the first layer of surgical mask followed by electron magnified images in \textbf{b-e}.
    \textbf{f}, digital image of the second layer of the surgical mask followed by electron magnified images in \textbf{g-j}.
    \textbf{k}, digital image of the third layer of the surgical mask followed by electron magnified images in \textbf{l-o}, \textbf{p}, nano-droplets on pristine silicon wafer piece collected in an open environment with a surgical mask and its magnified images in \textbf{q-t}.}
    \label{SI Figure:3}
\end{figure*}

The paint particles without any mask as a filter sprinkled on the silicon wafer piece are also shown in Figure \ref{SI Figure:3}.
The spray paint particles range from 80 nm to 40,000 nm which are mostly in circular shapes. 
The efficiency of filtration of two masks was calculated using nanoparticles of spray paint particles transmitted through different masks collected at a distance of 1 mm on a silicon wafer. 

In Figure \ref{SI Figure:3}\textbf{a-d}, the first layer of the mask has large numbers of nanoparticles on the surface. 
The second layer of the surgical mask, which is shown in Figure \ref{SI Figure:3}\textbf{f-j} has a fibre network to reduce the particle counts. 
In the final layer of the mask, Figure \ref{SI Figure:3}\textbf{k-o}, the particle counts were reduced, and the particle sizes of 80 nm -- 180 nm are found at the third layer.
We then collected the nanodroplets in pieces of the silicon wafer.
In Figure \ref{SI Figure:3}\textbf{p}, nano-droplets on pristine silicon wafer piece collected in an open environment with a surgical mask and its magnified images in Figure \ref{SI Figure:3}\textbf{q-t}.

\begin{figure*}[htp]
    \includegraphics[width=1\textwidth]{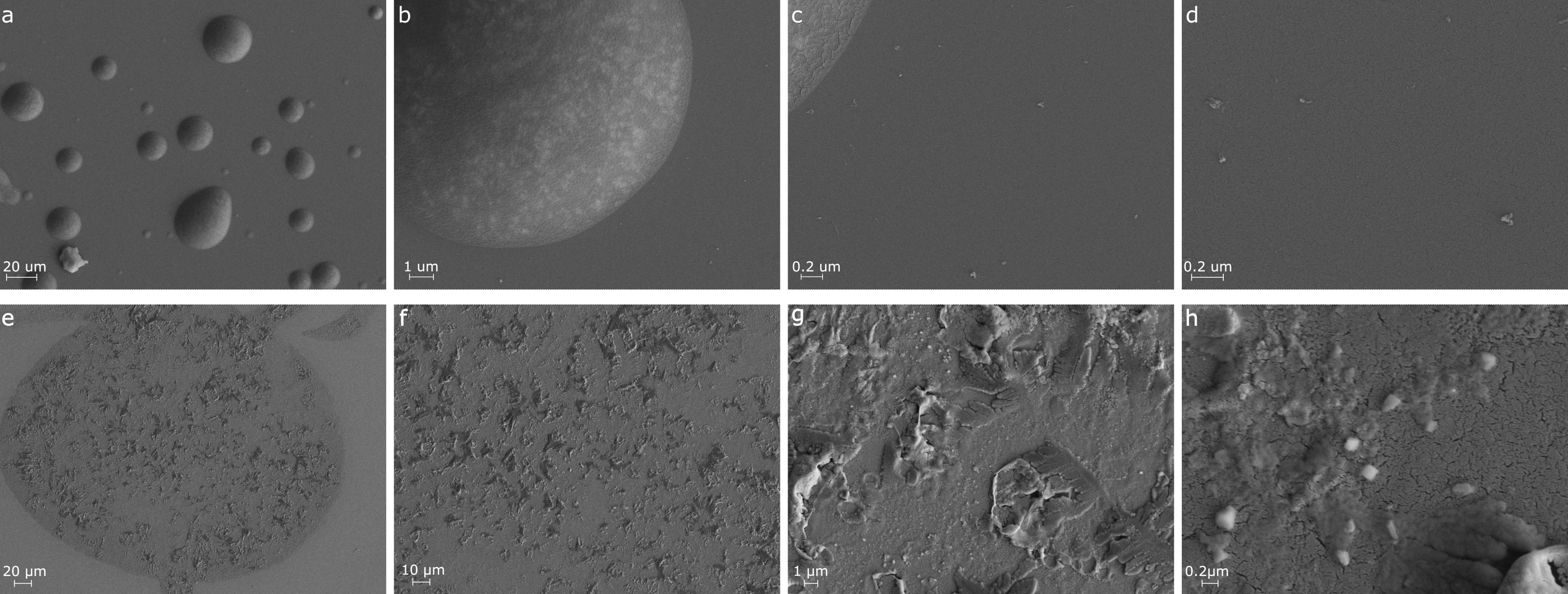}
    \caption{Spray paint nanoparticles on a silicon wafer and nanodroplets/particles collected in pristine silicon wafer pieces from spray paint and from an open environment.
    \textbf{a-d}, raw silicon wafer where spray paint particles are deposited on the surface of silicon without any mask as a barrier and its magnified images.
    \textbf{e-h}, nano-droplets on pristine silicon wafer pieces collected in an open environment without a mask as a barrier, and their magnified images were shown.}
    \label{SI Figure:4}
\end{figure*}

\begin{figure*}[htp]
    \centering
    \includegraphics[width=0.8\textwidth]{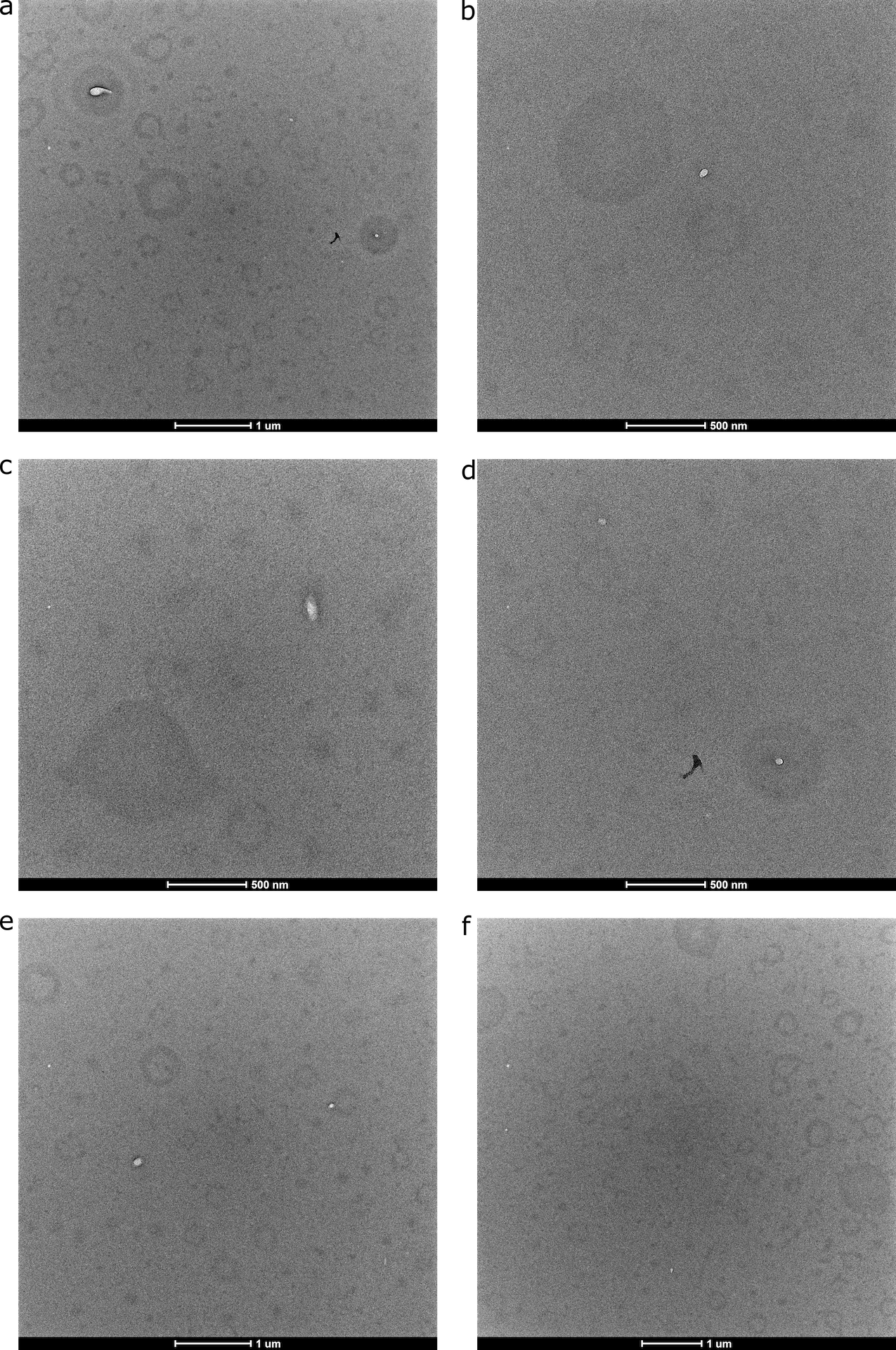}
    \caption{Transmission electron microscopy images of spray paint nanoparticles which are drop-casted on a carbon-coated copper grid.
    \textbf{a-f} shows the transmission electron microscopy images of spray paint nanoparticles which were distributed with low concentration throughout the grid. 
    Spray paint nanoparticles were in the range of 50 nm to 300 nm with different shapes.
    The background noise is the organic molecules i.e isopropyl alcohol used for the experiment}
    \label{SI Figure:5}
\end{figure*}

\begin{figure*}[htp]
    \centering
    \includegraphics[width=1\textwidth]{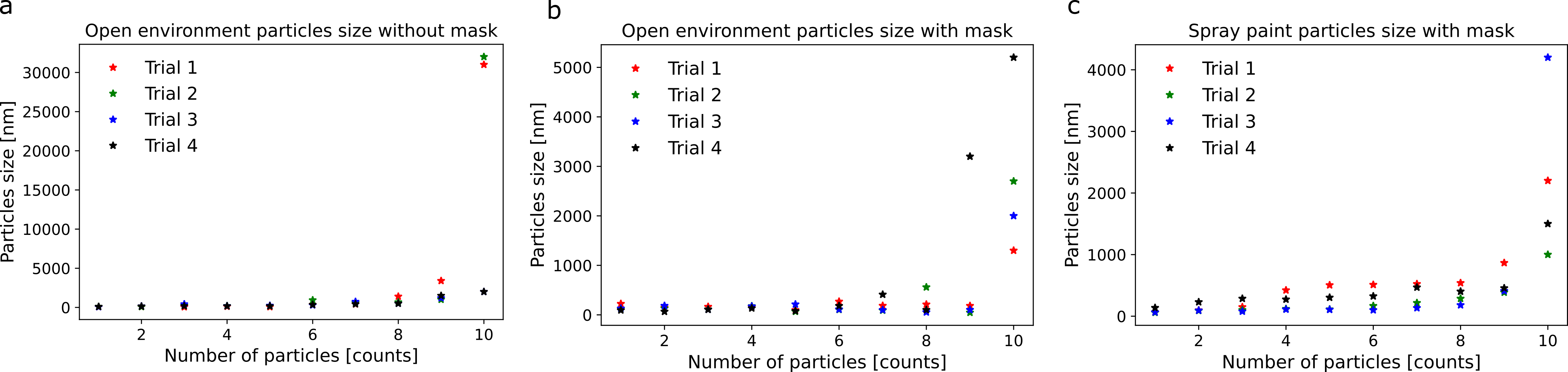}
    \caption{Reproducibility plot of the experiment.
    \textbf{a}, reproducibility plot of four trials of open environment particle size without a mask.
    \textbf{b}, reproducibility plot of four trials of open environment particle size with a mask.
    \textbf{c}, reproducibility plot of four trials of spray paint particle size with a mask.}
    \label{SI Figure:6}
\end{figure*}
\end{document}